\documentclass[11pt,eqsecnum]{article}

\usepackage{amsmath, amssymb}
 \usepackage{epsfig}
\usepackage{mathrsfs}
\usepackage{psfrag}
\usepackage{graphicx}
\usepackage{epstopdf}

\usepackage{color}

\newcommand \be{\begin{eqnarray}}
\newcommand \ee{\end{eqnarray}}

\newcommand{\ba}{\begin{eqnarray}}
\newcommand{\ea}{\end{eqnarray}}

\def\e{\epsilon}

\def\l {\lambda}

\def \td {\tilde}
\def \td {\tilde}
\def\sn{{\rm sn}}
\def\cn{{\rm cn}}
\def\dn{{\rm dn}}

\def \EE {{\mathbb{E}}}
\def \KK {{\mathbb{K}}}

\def\no{\nonumber}
\def\d{\partial}

\def \g{\gamma}
\def\o{\omega}
\def\S{{\cal S}}

\def\a{\alpha}
\def\k{\kappa}

\def\s{\sigma}

 \def\tg{\tilde{\g}}

\def \adss  {$AdS_5 \times S^5$}
\def \sql {\sqrt{\lambda}}
\def\O{{\cal O}}

\def\La{\Lambda}

\begin{document}
\renewcommand{\thefootnote}{\arabic{footnote}}
 
\def \foot {\footnote}
\def \bi{\bibitem}

\def \tr {{\rm tr}}
\def \ha {{1 \over 2}}

\def \ci{\cite}
\def \N {{\mathcal N}}
\def \const {{\rm const}}
\def \t {\tau}
\def\S{{\mathcal S} }
\def \nn {\nu}
\def \XX {{\rm X}}

 \def \vp {\varphi} \def \bs {\bar \s }
\def \k {\kappa}
\def\foot{\footnote}
\def \four{{\textstyle {1\ov 4}}}
 \def \third { \textstyle {1\ov 3
}}
\def\det{\hbox{det}}
\def \ci {\cite}
\def \ov {\over}

\def \bp {\begin{pmatrix}}  \def \epm {\end{pmatrix}}
\def \ha {{\textstyle{1 \ov 2}}}

\def \bi {\bibitem}
\def \la {\label}
\def \Tr  {{\rm Tr}}

\def \T {{\cal T}}
\def \l {\lambda}
\def\foot{\footnote}
\def \tl  {{\tilde \l}}
\def \sql {{\sqrt \l}}
\def \adss {$AdS_5 \times S^5$\ }
\newcommand{\rf}[1]{(\ref{#1})}

\def \bp {\begin{pmatrix}} 
 \def \emp {\end{pmatrix}}
 
 \def \dett  {{\det}}
 
\def \qr {{\hat \rho}}
\def \const {{\rm const}}
\def \bea{\begin{eqnarray}}
\def \eea{\end{eqnarray}}
\def \no {\nonumber}
\def \ov {\over}
\def \tr {{\rm Tr}}
\def \g {\gamma}
\def \tm {\mbb{T}}
\def \ha {\fr{ 1}{ 2}}
\def \half {\fr{ \trm{1}}{\trm{2}}}
\def \s {\sigma}
\def \vp {\varphi}
\def \td {\tilde}
\def \z {\zeta}
\def \H {{\rm H}} 
\def \Tr {{\rm Tr}}
\def \ep {\epsilon}
\def \bp {\begin{pmatrix}} 
 \def \emp {\end{pmatrix}}
\def \ef {\end{document}}
\def \del {\partial}
\def \G {\Gamma} \def \ha { { 1 \ov 2}}  \def \tg  {\td \Gamma} \def \m {\mu}
 \def \tdb {\bar } 
\def \lm {Lam\'e\ }
 \def \tdb {\bar } 

\overfullrule=0pt
\parskip=2pt
\parindent=12pt
\headheight=0in \headsep=0in \topmargin=0in \oddsidemargin=0in

\vspace{ -3cm} \thispagestyle{empty} \vspace{-1cm}
\begin{flushright} AEI-2010-150
\end{flushright}
\begin{center}
{\Large\bf
Quark-antiquark potential in $\mathbf{AdS}$ at one loop}

 \vspace{0.8cm} {
 Valentina~Forini \footnote{forini@aei.mpg.de} 
 }\\
\vskip  0.5cm

\small
{\em 
Max-Planck-Institut f\"ur Gravitationsphysik,  Albert-Einstein-Institut \\
Am M\"uhlenberg 1, D-14476 
Potsdam, Germany
}\normalsize
\end{center}

 \vskip 0.8cm

 \begin{abstract}
We derive an exact analytical expression for the one-loop partition function of a string in $AdS_5\times S^5$ background with world-surface ending on two anti-parallel lines. All quantum  fluctuations are shown to be governed by integrable, single-gap Lam\'e  operators. The first strong coupling correction to the quark-antiquark potential, as defined in $\mathcal{N}=4$ SYM, is derived as the sum of  known mathematical constants and a  one-dimensional integral representation. Its full numerical value can be given with arbitrary precision and confirms a previous result. 
 \end{abstract}


\renewcommand{\theequation}{1.\arabic{equation}}
 \setcounter{equation}{0}

\section{Overview}

The chance of studying weakly-coupled string theory to gain insight into strongly-coupled gauge theory, provided by the AdS/CFT correspondence, has a seminal example in the string realization of the quark-antiquark Wilson loop~\cite{Maldacena,Rey}, with heavy quarks modeled by W-bosons.  
The expectation value of the  rectangular  loop with length $T$ and width $L$, which in the limit $T\gg L$ can be seen as a pair of anti-parallel lines (the ``quark'' trajectories) at distance $L$, is  given by the effective energy of a string  on $AdS_5\times S^5$ whose ends, restricted to the four-dimensional boundary of $AdS_5$, are at a distance $L$ apart.  
In this context, the potential exhibits a  Coulomb-like law 
\be
V_{q\bar q}\,(\lambda,L)=-\frac{c(\lambda) }{L}~,
\ee
where $c(\l)$ is a function of the string tension (or 't Hooft coupling) that behaves as
 \be\la{c}
c(\lambda)=\begin{cases} 
\frac{\l}{4\,\pi}\,\Big[1-\frac{\lambda}{2\,\pi^2}\Big(\ln\frac{2\pi}{\lambda}-\gamma_E+1\Big)+{\cal O}(\lambda^2)\Big]  & ~~~~~~~~~~~~\lambda\ll 1~,\\
\frac{\sqrt{\lambda}\,\pi}{4\,\KK^2}\, \,\Big[1+\frac{a_1}{\sqrt{\lambda}}+{\cal O}\Big(\frac{1}{(\sqrt{\lambda})^2}\Big)\Big] &~~~~~~~~~~~~ \lambda\gg 1~.
\end{cases}
\ee
Above, the weak-coupling expansion  is the field-theoretical calculation of~\cite{ESZ, Pineda}, and $\KK=\KK(\textstyle{\frac{1}{2}})$ is the complete elliptic integral of the first kind with modulus $k=\textstyle{\frac{1}{\sqrt{2}}}$~\footnote{See Appendix B for notation. We adopt here the Abramowitz-Mathematica notation for the modulus of the elliptic functions.}. 
The problem of finding the first quantum string correction  $a_1$ to the classical result of~\cite{Maldacena}\cite{ Rey}, initiated in~\cite{Kallosh,Greensite,Kinar},  has been first addressed in~\cite{Theisen,dgt}, resulting in a formal expression for the one-loop contribution to the effective action as a ratio of determinants of two-dimensional generalized Laplace operators. A numerical prediction for $a_1$ has been presented in~\cite{Chu}. Our main motivation here is to address the issue of exploiting exact analytical methods for computing the determinants in the partition function of~\cite{Theisen, dgt} and thus the analytically exact value of the constant $a_1$ in the sub-leading correction to the  potential. 

The evaluation of quantum corrections to the  energies of classical string solutions  in $AdS_5\times S^5$ \cite{reviewsemiclassical}, crucial device for checking the detailed structure as well as the integrability of the AdS/CFT system~\cite{reviewintegrability,Didina},  is in 
general a hard mathematical problem. The task is  simplified considering  scaling limits of some ``semiclassical parameters'', as in the case of fluctuations over the open string solution dual to the cusp Wilson loop~\cite{Kruczenski:2007cy,Roiban:2007ju,Roiban:2007dq},  or the closed string solutions of~\cite{ft1, Frolov:2003tu,ptt,ftt}. In these  limits  the solutions become linear in the world-sheet coordinates $(\tau,\sigma)$, thus making constant the coefficients in the fluctuation Lagrangian.  In the case of the Wilson loop of a pair of anti-parallel lines,  which has no other parameters than the distance between the lines,  the complicated $\s$-dependence of the lagrangean coefficients makes non-trivial  the  evaluation of the operator spectra. The same is true for the straight line and circular Wilson loops~\cite{Theisen,dgt,Zarembo:2002an}, for which a first explicit computation of  fluctuation determinants has been carried out in~\cite{KT}. There, based on the effective one-dimensionality of the spectral problem, 
it was possible to trade the explicit evaluation of the eigenvalue spectrum for the relevant operator with the resolution of the associated differential equation, an approach known as  Gelfand-Yaglom method~\cite{GelfandYaglom,dunne}.  In an analogous fashion  the case of the anti-parallel lines has been studied in~\cite{Chu}, where each functional determinant has been formally expressed in terms of the  associated initial value problem with Dirichlet boundary conditions (the appropriate ones in this framework~\cite{DGO}). While the possibility of a completely analytical treatment of such initial value problem was not recognized in~\cite{Chu},   the coefficient $a_1$ in (\ref{c}) was worked out by the authors with great numerical precision. 
A step forward in the exact analytical treatment of string quantum corrections has been made in~\cite{BDFPT} for the case of the folded string solution~\cite{gkp,ft1}, and recently in~\cite{others} for the case of pulsating strings. It has been there realized that  fluctuations on this basic class of elliptic solutions can be put into the standard (single-gap) Lam\'e form, which allows an  exact treatment  of the fluctuation problem.  This is useful to extract information in the meaningful semiclassical limits of large~\cite{bftt,reviewreciprocity} and short values of the conserved charges typical of the problem~\cite{tt,rt,Basar}.

We revisit here the evaluation of the one-loop partition function that defines the first subleading correction to the quark-antiquark  $AdS$ potential,  showing that also in this case the fluctuations are governed by Lam\'e operators. This allows us to present some analytically exact results, as the expressions for the fluctuation determinants (\ref{detO1})-(\ref{detOf}) with (\ref{alpha1}), (\ref{alpha2})-(\ref{sigma}) and the resulting formula for the one-loop partition function  (\ref{Van})-(\ref{integral}) with (\ref{integral3})-(\ref{x2}), see also the equivalent expressions collected in Appendix D.   Finally, we find for $a_1$ the following representation
\ba\la{constant}
a_1&=&\frac{5 \pi }{12}-3 \ln 2+\frac{2 \KK}{\pi}\,\big(\,\KK-\sqrt{2}\,(\pi+\ln2)+{\cal I}^{\rm num}\,\big)\\\no
&=&-1.33459530528060077364\dots~,
\ea
where  the contribution ${\cal I}^{\rm num}$, whose one-dimensional integral representation is displayed in (\ref{numerics}), can be evaluated with arbitrary precision. The numerical value of $a_1$  confirms the result obtained in~\cite{Chu}.

The connection of the fluctuation problem to the integrable Lam\'e differential equation is not surprising, since the minimal surface corresponding to the  Wilson loop of anti-parallel lines belongs, as the folded and pulsating string cases,  to the important class of classical string solutions  expressed in terms of elliptic functions (see Appendix A). 
It is however interesting to see on this non-trivial example how the integrability of the $\sigma$-model on $AdS_5\times S^5$~\cite{bpr}\cite{kaz,gr1} is extended from the classical to the one-loop level via this special, integrable, type of potential. It is also interesting to recall that the chance of exploiting the integrability of the underlying sigma-model to calculate  Wilson loops within the AdS/CFT correspondence~\cite{DrukkerFiol}  has been made recently concrete, via the connection of Wilson loops to $\mathcal{N}=4$ SYM scattering amplitudes,  with the proposal of Thermodynamic Bethe Ansatz equations for the latter~\cite{Alday:2010vh}. Although the Wilson loops appearing in amplitude computations consist of light-like segments that  are not obviously related to the configuration of space-like anti-parallel lines of interest here, it is fascinating to think about the possibility (on the lines of the approximation in~\cite{AldayComments}) of using a  description  similar to the one in~\cite{Alday:2010vh} also in this case.

It would be also interesting to exploit similar analytical methods in the case of the one-loop partition function for the anti-parallel lines configuration in a Schwarzschild-$AdS_5$ background~\cite{Greensite,Naik}, whose formal expression has been worked out in~\cite{Hou}.

The main body of this article contains the analytical study of the fluctuations  and the evaluation of the one-loop contribution to the quark-antiquark potential. Appendices A, B and C recall basic facts on the world-sheet set-up, elliptic integrals and functions and on the Gelfand-Yaglom method. Alternative expressions for the relevant integrals are displayed in Appendix D.

\renewcommand{\theequation}{2.\arabic{equation}}
 \setcounter{equation}{0}
 \section{Fluctuation operators and their analytical determinants}

Given the invariance of the anti-parallel lines configuration under time-translation, both the bosonic and the fermionic fluctuation lagrangeans depend non-trivially only on the $\sigma$-coordinate, and the original two-dimensional spectral problem is reduced to the evaluation of \emph{one-dimensional} functional determinants.
After suitable world-sheet reparametrization and fermion diagonalization~\cite{dgt,Chu} reviewed in Appendix A, the resulting effective action for a string in $AdS_5\times S^5$ background with world-surface ending on two anti-parallel lines~\cite{Theisen, dgt}
can be written as follows upon Fourier transformation of the time variable  ($\partial_\tau=-i\,\omega$)
\be\la{Gammapar}
\Gamma_{||}=-{\cal T}\,\int\,\frac{d\omega}{2\pi}\ln  \frac{\det^{2}{\cal O}_+\,\det^{2}{\cal O}_- }{\det  {\cal O}_1\,\det^{1/2}{\cal O}_2\,\det^{5/2}{\cal O}_0}~,
\ee
where ${\cal T}=\int d\tau$ is the $\tau$-period. Above,  ${\cal O}_0=-\d^2_\s +\o^2$ is the free operator and
\be \la{eq}
 \O_i  = -\d^2_\s+ V_i(\s) +\o^2\,~~~~~~~~~~~~i=1,2,\pm ~,
 \ee
 \be\la{V}
V_1= \frac{1}{\cn^2\s}\ , ~~~~~~~~~~~~
V_2  =  \frac{1}{\cn^2\s}-\cn^2\s , ~~~~~~~~~~~~
V_{\pm} = \frac{1\pm\sqrt{2}\,\sn \s\,\dn\s}{2\,\cn^2\s}~.
\ee
The Jacobi elliptic functions appearing in (\ref{V}) and defined in Appendix B have fixed modulus $k=\frac{1}{\sqrt{2}}$ and  $-\KK<\s<\KK$. The operators $ \O_i $ are also defined in (\ref{opbos0})-(\ref{diag}).

The partition function (\ref{Gammapar}) suffers in general from linear infrared divergencies, that can be cured subtracting a reference solution (as in~\cite{KT} for the example of the circular Wilson loop). The one-loop correction to the quark-antiquark potential can be  thus obtained  subtracting twice the infinite, self-energy contribution of each of the parallel lines (quarks)~\cite{Chu}, and dividing over the infinite time period  $T=\int dt$
\ba\label{qqformal}
V^{(1)}_{q\bar q}=\lim_{T\to\infty}\frac{1}{T}\,\Big[\Gamma_{||}-2\,\Gamma_{|}\Big]~,~~~~~~~~~~~~~~T=\frac{\KK\,L}{\pi}\,{\cal T}\to\infty~,
\ea
where the relation between $T$ and ${\cal T}$ follows from  (\ref{y0}) and (\ref{reparam})~\footnote{The world-sheet of the straight line can be parametrized with the same time variable as the one  for the anti-parallel lines~\cite{Chu}, see (\ref{reparam}).}.
   
Exploiting elementary transformations of Jacobian elliptic functions~\cite{Abram}, it is easy to check that  each non-trivial fluctuation  operator  is a  \emph{single-gap Lam\'e operator} with  the following  eigenvalue problem  
\be\label{lamegen}
\Big[-\d^2_x+\sn^2x+\Omega^2\,\Big]\,f_\Lambda(x)=\La\,f_\La(x)\ ,
\ee
where, with respect to (\ref{eq}),   $x$ and $\Omega$ are a shifted (and rescaled) $\s$ variable and euclidean frequency respectively. Explicitly, 
\begin{description}
\item{(a)} for the first bosonic operator $V_1$:~ $x=\s+(1+i)\,\KK$ and ${\Omega}^2 =\omega^2-1$;
\item{(b)} for the second bosonic operator  $V_2$: $x=(1+i)\,\s+\KK$ and ${\Omega}^2 =\frac{\omega^2}{2i}-1$;
\item{(c)} for the fermionic  operators $V_\mp$:  
$x=\begin{cases} 
\frac{\s}{2}(-1+i)+\frac{\KK}{2}(1+i),  & \mbox{for}~ V_-\\
\frac{\s}{2}(-1+i)+\frac{\KK}{2}(3-i), &  \mbox{for}~ V_+
\end{cases}
 ~,  ~{\rm and} ~~~\Omega^2=2\,i\,\omega^2-1~.
 $
\end{description}
The Lam\'e  spectral problem  (\ref{lamegen})  can be solved \emph{exactly}, and hence the corresponding determinant  can be  computed \emph{analytically}, relying on the knowledge of the solutions to (\ref{lamegen}) and the use of the Gelfand-Yaglom method (see, for example,~\cite{dunne} and~\cite{BDFPT}\cite{others}). While the general procedure  is briefly reviewed in Appendix C, let us see explicitly the evaluation of the determinant for the bosonic fluctuation governed by the first potential $V_1$.

\bigskip

Readapting the solutions (\ref{solslame}) to the case (a) above, two independent solutions of the relevant differential equation are 
\be\la{solutionsV1}
y_\pm(\s)=\frac{H(\s+\KK(1+i)\pm\alpha_1)}{\Theta(\s+\KK(1+i))}\,e^{\mp Z(\alpha_1)\,(\s+\KK(1+i))}\equiv\,
\frac{\theta_3\big(\frac{\pi(\s\pm\alpha_1)}{2\KK}\big)}{\theta_2\big(\frac{\pi\,\s}{2\KK}\big)}\,e^{\mp Z(\alpha_1)\,(\s+\KK(1+i))}~,
\ee
where the Jacobi $H$, $\Theta$ and $Z$ functions are defined in (\ref{jacobidef}) in terms of the Jacobi $\theta$-functions and
\be\la{alpha1}
\alpha_1=\sn^{-1}\sqrt{\frac{k^2+\o^2}{k^2}}\equiv\sn^{-1}\sqrt{1+2\,\o^2}~.
\ee
The solutions (\ref{solutionsV1}) \emph{diverge} at the extrema $\s=-\KK$ and $\s=+\KK$ of the interval, which is  a direct way to see the standard need~\cite{DGO} of an infrared regulator $\epsilon$. The Gelfand-Yaglom theorem will be therefore applied solving the initial value problem in the interval $-\KK+\epsilon<\s<\KK-\epsilon$ where $\epsilon$ is arbitrary small. 
The linear combination 
\be\la{u2}
u(x)=\frac{y_+(-\KK+\epsilon)\,y_-(x)-y_-(-\KK+\epsilon)\,y_+(x)}{W(-\KK+\epsilon)}~,
\ee
with wronskian $W(x)$, see (\ref{wronskian}), evaluated at the regularized initial point, is a solution of the homogeneous equation with boundary conditions
\be
u(-\KK+\e)=0,~~~~~~~~~u'(-\KK+\e)=1~.
\ee
As follows from the discussion in Appendix C, the determinant of the bosonic operator $V_1$ with Dirichlet boundary conditions in the interval $[-\KK+\epsilon,\KK+\epsilon]$ will be then given by $u(\KK-\e)$. One finds 
\be\la{detO1}
\det{\cal O}_1=\frac{2\,{\rm ns}^2\e-\sn^2\a_1}{\sn\a_1\,\cn\a_1\,\dn\a_1}\,\sinh[\,2\,Z(\a_1)\,(\KK-\e)+\Sigma_1\,],~~~~~~~~~~\Sigma_1=\ln\frac{\theta_4\big(\frac{\pi \,(\alpha_1+\e)}{2\KK}\big)}{\theta_4\big(\frac{\pi \,(\alpha_1-\e)}{2\KK}\big)}~.
\ee

\bigskip
In a similar fashion one can work out the regularized determinants for the potential $V_2$ and $V_\pm$, obtaining 
\ba\la{detO2}
\!\!\!\!\!\!\!\!\det{\cal O}_2&=&\!\!\!\!\frac{1}{(1+i)}\frac{2\,{\rm ns}^2[(1+i)\,\e]-\sn^2\a_2}{\sn\a_2\,\cn\a_2\,\dn\a_2} 
\,\sinh\Big[\,2\,Z(\a_2)\,(1+i)(\KK-\e)+\frac{i\,\pi\,\alpha_2}{\KK}+\Sigma_2\,\Big],~~~~~~\\\la{detOf}
\!\!\!\!\!\!\!\!\det{\cal O}_f&=&\!\!\!\!(1+i)\frac{2\,{\rm dn}^2\a_f-{\rm nd}^2\big[\frac{\e}{(1+i)}\big]}{\sn\a_f\,\cn\a_f\,\dn\a_f}\,\frac{\theta_3\big(\frac{\pi}{2\KK}\frac{\e}{1+i}\big)}{\theta_1\big(\frac{\pi}{2\KK}\frac{\e}{1+i}\big)}\,\sinh\Big[\,2\,Z(\a_f)\,\frac{(\KK-\e)}{1+i}-\frac{i\,\pi\,\alpha_f}{2\,\KK}+\Sigma_f\,\Big],
\ea
where
\ba\la{alpha2}
\alpha_2=\sn^{-1}\sqrt{1-i\,\o^2}~,~~~~~~~~~
\alpha_f=\sn^{-1}\sqrt{1+4i\,\o^2}~,
\ea
and
\be\la{sigma}
\Sigma_2=\ln\frac{\theta_4\big(\frac{\pi \,(\alpha_2+(1+i)\e)}{2\KK}\big)}{\theta_4\big(\frac{\pi \,
(\alpha_2-(1+i)\e)}{2\KK}\big)}~,~~~~~~~~~~
\Sigma_f=\ln\frac{\theta_4\big(\frac{\pi}{2\KK} \,\big(\alpha_f+\frac{\e}{1+i})\big)}{\theta_2\big(\frac{\pi}{2\KK}\big(\alpha_f-\frac{\e}{1+i})\big)\big)}~.
\ee
Notice that in the fermionic case  $\det\,{\cal O}_+\!=\det\,{\cal O}_-\!\equiv\det\,{\cal O}_f$. This can be understood by noticing, in (\ref{V}), that $V_+(-\sigma)=V_-(\sigma)$. Namely, it holds that ${\cal O}_+={\cal P}^{-1}\,{\cal O}_-\,{\cal P}$, with ${\cal P}$ the unitary parity operator with respect to $\sigma$, implying  $\det\,{{\cal O}_+}\!\equiv \det\,{\cal O}_-\!$~\footnote{In this case, in which the evaluation of determinants is done via the Gelfand-Yaglom theorem, the equivalence of the determinants  can be easily checked  exploiting this parity feature as inherited in the solutions via which the determinant is defined, see (\ref{u2}) and (\ref{wronskian}).}.

The contribution of the massless bosons can be easily obtained via the same method 
\be
\det{\cal O}_{0}=\frac{\sinh[2\omega(\KK-\epsilon)]}{\omega}.
\ee 
 
Expanding in $\e\sim0$ and retaining the divergent contributions, one gets
\ba\la{O1epsilon}
\det{\cal O}_1^\epsilon&\cong&-\frac{2}{\epsilon ^2\, \omega ^2}\,\sqrt{\frac{\omega ^2}{4\, \omega ^4-1}} \,\sinh[\,2 \,\KK\, Z(\a_1)\,]~,\\\la{O2epsilon}
\!\!\!\!\!\!\!\!\!\!\!\!\!\det{\cal O}_2^\epsilon&\cong&-\frac{1}{\epsilon ^2\, \omega ^2}\, \sqrt{\frac{\omega ^2}{\omega ^4+1}} \,\sinh \Big[\,2\,(1+ i)\, \KK\, Z(\alpha_2)+\frac{i\, \pi\, \alpha_2}{\KK}\,\Big]~,\\\la{Ofepsilon}
\!\!\!\!\!\!\!\!\!\!\!\!\!\det{\cal O}_f^\epsilon&\cong&\frac{4}{\e}\frac{1}{  \sqrt{16
  \, \omega ^4+1}} \sin\,\Big[(1+i) \,\KK\, Z(\alpha_f)+\frac{\pi \,\alpha_f}{2 \,\KK}\Big]~,\\\la{O0epsilon}
\!\!\!\!\!\!\!\!\!\!\!\!\!\det{\cal O}_0^\epsilon&\cong& \frac{\sinh[\,2\,\KK\, \o\,]}{\o}~.
\ea
As usual, the divergence $\sim1/\epsilon$ in the resulting ratio of determinants as they appear in (\ref{Gammapar}) is cured subtracting twice the contribution of the straight line, which can be evaluated and regularized by the same means~\cite{KT}\cite{Chu}
\be\la{Gammastraight}
\Gamma_{|}^\epsilon=-\frac{\cal{T}}{2}\int_{-\infty}^\infty\,\frac{d\omega}{2\pi}\,\ln \Big[1+\frac{1}{\o\,\epsilon}\Big]~.
\ee

\renewcommand{\theequation}{3.\arabic{equation}}
 \setcounter{equation}{0}
\section{ One-loop correction to the quark-antiquark potential}

The one-loop correction to the quark-antiquark potential is formally defined by (\ref{qqformal}), in which $\Gamma_{||}$ is given in terms of the determinants (\ref{O1epsilon})-(\ref{O0epsilon}),  $\Gamma_{|}$ is substituted by the regularized expression  (\ref{Gammastraight}) and the regulator $\e$ is sent to zero.  Namely, it is
\ba\la{V1def}
\!\!\!\!\!\!\!\!V^{(1)}_{q\bar q}=-\frac{1}{2}\,\frac{\pi}{\KK\,L}\lim_{\epsilon\to0}\int_{-\infty}^{+\infty}\frac{d\omega}{2\pi}\left[\ln\Big[\frac{\det^{8}{\cal O}_f ^\epsilon}{(\det  {\cal O}_1^\epsilon)^2\,\det{\cal O}_2^\epsilon\,\det^{5}{\cal O}_0^\epsilon}\Big]-2\ln\Big[1+\frac{1}{\o\,\epsilon}\Big]\right]~.
\ea
Making (\ref{V1def}) explicit in terms of the determinants (\ref{O1epsilon})-(\ref{O0epsilon}), the following exact \emph{analytical} expression is obtained for the one-loop correction to the quark-antiquark $AdS$ potential 
\ba\la{Van}
\!\!\!\!\!\!\!\!V^{(1)}_{q\bar q}&=&-\frac{1}{2\,\KK\,L}\int_{0}^{+\infty} d\omega
\ln\Big[\frac{128^2 \omega ^{10} \left(1-4 \omega ^4\right)\sqrt{1+\omega^4}}{\left(16 \omega ^4+1\right)^4}\Big]+\\\no
\!\!\!\!\!\!\!\!&&-\frac{1}{2\,\KK\,L}\int_{0}^{+\infty}d\omega
\ln\left[\frac{ \sin
   ^8\Big[(1+i)\KK\, Z(\alpha_f)\,+\frac{\pi  \alpha_ f}{2 \KK}\Big]}{\sinh^2[2\,\KK\,Z(\a_1)]\,\sinh \Big[2(1+
   i)\KK\,Z(\alpha_2)+\frac{i \pi \alpha_2}{\KK}\Big]\,\sinh^{5}[2\,\KK\,\o]}\right]\\\la{integral}
   &=&\frac{\pi}{\sqrt{2}\,\KK\,L} -\frac{1}{2\,\KK\,L}\,{\cal I}~.
\ea
Above, $\a_1,\a_2,\a_f$ are implicitly defined in (\ref{alpha1}), (\ref{alpha2}) and in the last equivalence we have  reported the result for the integral in the first line.
The second non trivial integral $\cal I$ can be partially given in terms of known mathematical constants. One can proceed rewriting it as~\footnote{A quick way to obtain this expression consists in taking the derivative of the arguments of the hyperbolic functions in (\ref{integral}) and integrating back using standard tables of integrals.}
\be\la{integral3}
{\cal I}=\!\!\!\int_{0}^{k}\!\!\! d\omega \,\ln\Big[\frac{\cosh^8 x_f}{\cos^2  x_{1{\rm zero}}\,\sinh x_2\,\sinh^5[2\,\KK\,\o]}\Big]+\int_{k}^{\infty}\!\!\!d\omega \,\ln\Big[\frac{\cosh^8  x_f}{\sinh^2 x _{1{\rm inf}}\,\sinh x_2\,\sinh^5[2\,\KK\,\o]}\Big] 
\ee
where 
\ba\la{xf}
&&\!\!\!\! \!\!\!\!\!\!\!\! \!\!\!\! \!\!\!\! 
 x_f=\KK\, \Big(\,\textstyle{\frac{1}{2}}F\big[\cos^{-1}(\textstyle{\frac{1-4 \omega ^2}{1+4 \omega ^2}})\big]-E\big[\cos^{-1} (\textstyle{\frac{1-4 \omega ^2}{1+4 \omega ^2}})\big]+\textstyle{\frac{2\o\, \sqrt{16 \omega ^4+1} }{1+4 \omega ^2}}\,\Big)+\frac{\pi}{4\,\KK}F \big[\cos^{-1} (\textstyle{\frac{1-4 \omega ^2}{1+4 \omega ^2}})\big]~,\\\la{x1zero}
&&\!\!\!\!\!\! \!\!\!\!\!\!\!\! \!\!\!\! \!\!\!\! \!\!\!\!\!\!
x_{1{\rm zero}}=\KK\,
   \left(\,F\big[\cos^{-1}(\sqrt{2} \omega )\big]-2 E\big[\cos^{-1}(\sqrt{2} \omega)\big]\,\right)+\textstyle{\frac{\pi}{2\,\KK} } F\big[\cos^{-1}(\textstyle{\frac{\sqrt{1-2 \omega ^2}}{\sqrt{1+2 \omega ^2}}})\big]~,\\\la{x1inf}
&&\!\!\!\!\! \!\!\!\!\!\!\!\! \!\!\!\! \!\!\!\! \!\!\!\!
x_{1{\rm inf}}=\KK\,\Big(\,F\big[\sec^{-1}(\sqrt{2}\,\o)\big]
-2 E\big[\sec^{-1}(\sqrt{2}\,\o)\big]
   +\textstyle{\frac{\sqrt{4 \omega ^4-1}}{\omega }}\Big)
   +\frac{\pi }{2}-\frac{\pi}{2\,\KK}  F\big[\sec^{-1}(\textstyle{\frac{\sqrt{2 \omega ^2+1}}{\sqrt{2 \omega
   ^2-1}}})\big]~,\\\la{x2}
&&\!\!\!\! \!\!\!\!\!\!\!\! \!\!\!\! \!\!\!\! 
x_2=\KK\, \Big(\,F\big[\cos
   ^{-1}(\textstyle{\frac{1-\omega ^2}{1+\omega ^2}}\big)-2 \,E\big[\cos ^{-1}(\textstyle{\frac{1-\omega ^2}{1+\omega
   ^2}})\big]+\textstyle{\frac{2  \omega\,\sqrt{1+\omega ^4} }{\omega ^2+1}}\,\Big)+\frac{\pi}{2 \,\KK}  F\big[\cos ^{-1}(\textstyle{\frac{1-\omega ^2}{1+\omega ^2}})\big]~,
  \ea
  and  $E[x]\equiv E[x,\textstyle{\frac{1}{2}}]$ and $F[x]\equiv F[x,\textstyle{\frac{1}{2}}]$ are
 the incomplete elliptic integrals of the first kind defined in (\ref{incomplete}). The need of  two separate intervals of integration, $\o\in(0,k)$ and $\o\in(k,\infty)$, is due to the bosonic fluctuations described by ${\cal O}_1$ and is clear, for example, by looking at the square root in (\ref{O1epsilon}).
  
Rewriting now
\be\la{logsinh}
\ln\sinh x=\ln\textstyle{\frac{1}{2}}+x+\ln[1-e^{-2x}],
\ee 
its analogue for the $\cosh$, and disregarding the constant contribution which will vanish due to the balance of the world-sheet degrees of freedom, one can consider the  part $\sim x$ in (\ref{logsinh}) and work out some analytical finite contribution (the logarithmical and power-like divergencies  will cancel in the ratio).  The numerical integration for the remaining contribution $\sim\ln[1-e^{-2x}]$  converges quickly to a steady value, and can be obtained via standard packages like {\rm Mathematica} with an \emph{arbitrary} precision. This way the contribution of each fluctuation can be evaluated separately. 

For example, in the case of the fermions the indefinite integration will give
\be
8\, \int_0^\omega\! d\omega'\, x_f=8\,\o^2\,\KK-\frac{2\,\pi}{\KK}\,\ln\o+2\KK-\frac{\pi}{\KK}  (2+3 \ln2)+{\cal O}\Big(\frac{1}{\o^4}\Big) ~.
\ee
In an analogous way one can evaluate the analytical contributions for all the fluctuations, check the cancellation of the divergent pieces and get for the finite ones 
\ba
{\cal I}^{\rm an}_{ferm}&=& 2\KK-\frac{\pi}{\KK}  (2+3 \ln2)~,~~~~~~~~~~~~~~~~~~~~
{\cal I}_{\rm free}=\frac{5\,\pi^2}{24\,\KK}~,\\
{\cal I}^{\rm an}_{1}&=&\frac{\pi}{\KK}(1+\ln2)-\sqrt{2}\,\ln2~,~~~~~~~~~~~~~~~~
{\cal I}^{\rm an}_{2}=\frac{\pi}{\KK}\,(1+\frac{\ln2}{2})-\KK~,
\ea
where ${\cal I}_{\rm free}$ amounts for the total contribution of the free fluctuations to ${\cal I}$.
 
The remaining contributions can be evaluated numerically with arbitrary precision~\footnote{Notice that the first term in ${\cal I}_1^{\rm num}$ automatically includes the type of constant contribution, $\ln\frac{1}{2}$, which should cancel in the balance of degrees of freedom. Such contribution, amounting  to 
$2 \int_0^k d\o \ln\textstyle{\frac{1}{2}}=-\sqrt{2}\,\ln2$, has then to be subtracted and is in fact  included in ${\cal I}^{an}_1$.}
\ba\la{numf}
{\cal I}^{\rm num}_{ ferm}&=&8\,\int_0^\infty d\o\,\ln[1+e^{-2\,x_f}]=1.41586~,\\\la{num1}
{\cal I}_1^{\rm num}&=&-2\int_0^k d\o\,\ln[\cos x_{1{\rm zero}}]-2\int_k^\infty d\o\,\ln[1-\exp(-2\,x_{1{\rm inf}})]=1.18174~,\\\la{num2}
{\cal I}^{\rm num}_{2}&=&-\,\int_0^\infty d\o\,\ln[1-e^{-2\,x_2}]=0.43859~.
\ea
Adding together the analytical and numerical contributions
\be\la{anandnum}
{\cal I}^{\rm an}= \frac{\pi}{\KK} \Big(\frac{5\,\pi}{ 24}-\frac{ 3}{2} \ln2\Big) +\KK -\sqrt{2}\,\ln2,~~~~~~~~~~~~~~~{\cal I}^{\rm num}=3.09111~,
\ee
it follows for the one-loop correction  in (\ref{integral}) the expression
\ba\la{halfanalit}
V^{(1)}_{q\bar q}=-\frac{1}{2\,\KK\, L}\,\Big[  \frac{\pi}{\KK} \Big(\frac{5\,\pi}{ 24}-\frac{ 3}{2} \ln2\Big)+\KK-\sqrt{2}(\pi+\ln2) +{\cal I}^{\rm num}\,\Big]=\frac{0.30492}{L}~.
\ea
A compact way to define ${\cal I}^{\rm num}$, which is equivalent to the sum of the contributions in (\ref{numf})-(\ref{num2}),  is given in  (\ref{numerics})  and evaluated there with high precision.
The $AdS$ quark-antiquark potential is therefore given by
\be\la{classplusoneloop}
V_{q\bar q}(\sqrt{\lambda},L)=-\frac{\sqrt{\lambda}\,\pi}{4\,\KK^2\,L}\,\Big[1+\frac{a_1}{\sqrt{\lambda}}+{\cal O}\Big(\frac{1}{(\sqrt{\lambda})^2}\Big)\Big]~,~~~~~~\l\gg1,
\ee
where  the one-loop correction $a_1$  is given in (\ref{constant}), and 
confirms the result obtained in~\cite{Chu}. As observed there, it is interesting to notice that, when compared  to the strong coupling prediction via summation of ladder diagrams of~\cite{ESZ}
\be
V^{\rm ESZ}_{q\bar q}(\sqrt{\lambda},L)=-\frac{\sqrt{\lambda}}{\pi\,L}\,\Big[1-\frac{\pi}{\sqrt{\lambda}}+{\cal O}\Big(\frac{1}{(\sqrt{\lambda})^2}\Big)\Big]~,~~~~~~\l\gg1,
\ee
$a_1$ has the same sign and smaller absolute value. 
 
To summarize, rephrasing the fluctuations over the minimal surface related to Wilson loop with anti-parallel lines in terms of the \lm spectral problem (\ref{lamegen}), we were able to present useful analytical formulas for the fluctuation determinants (\ref{detO1})-(\ref{detOf}), for the partition function  (\ref{Van})-(\ref{integral}), and finally the representation (\ref{constant}) for the first subleading correction to the quark-antiquark potential.



\section*{Acknowledgments }
I am grateful to  A. Tseytlin for bringing my attention to Ref.~\cite{Chu}, and I thank him and N. Drukker for useful comments on the draft. I am indebted to D. Seminara for valuable discussions and comments on the draft.  I also thank M. Beccaria,  V. Giangreco Marotta Puletti, A. Gorsky, N. Kim, T. Mc Loughlin, J. Russo and S. Theisen  for discussions.
During the work I enjoyed very kind hospitality at the University of Barcelona, at the Nordita Institute in Stockholm (IGST2010 conference and related  workshop) and at the Kyung Hee University and KIAS in Seoul.


 \bigskip
 

 \appendix
 \section*{Appendix A:  World-sheet set-up}
\refstepcounter{section}
\def\theequation{A.\arabic{equation}}
\setcounter{equation}{0}

At the classical level, the quark-antiquark potential in $AdS_5\times S^5$ described by the metric
 \be\la{metric}
 ds^2=y^2\,(dx^n)^2+\frac{dy^2}{y^2}+d\Omega^2_5~,
 \ee
  is evaluated considering two anti-parallel lines extended in the $x^0$ direction and located at $x^1=\pm\frac{L}{2}$. In (\ref{metric}), we set to $1$ the radius of both $AdS_5$ and $S^5$, $n=0,1,2,3$ and the index $4$ labels the coordinate $y$, ranging from its minimal value in the bulk to an infinite value on the boundary.
The world-sheet in the bulk can be parametrized by $(\tau,\s)=(x^0,x^1)$, $-{\cal T}/2<\tau<{\cal T}/2$. Given the invariance of the problem under $x^0$-translation, the surface stretched between the lines is a function $y=y(x^1)$ of the $x^1$ coordinate only, and the induced metric reads
\be
ds^2=y^2\,dt^2+\frac{1}{y^2}(y^4+y'^2)\,{d\s}^2.
\ee
Given the Nambu-Goto action
\be\la{NGaction}
S=\frac{\sqrt{\lambda}\,T}{2\pi}\int d\sigma\sqrt{y'^2+y^4}~,
\ee 
its equation of motion $y\,y''=4y'^2+2y^4$ has a first integral of motion 
 \be\la{solution}
 y'^2=\frac{y^8}{y_0^4}-y^4,
 \ee
 which can be can be integrated in terms of elliptic functions.
Above, $y_0$ is an integration constant corresponding to the minimal value of the coordinate $y$ in the bulk and is related to the distance $L$ between the lines via 
\be\la{y0}
y_0=\frac{\pi}{\sqrt{2}\,\KK\,L}. 
\ee
 One then proceeds evaluating the action (\ref{NGaction}) on the solution (\ref{solution})
 \be\la{classicalpot}
 S=\frac{\sqrt{\lambda}\,T}{2\pi\,y_0^2}\int_{-L/2}^{L/2}d\s\,y^4~~~~~~~~~\longrightarrow~~~~~~~~~S=-\frac{\sqrt{\lambda}\,\pi}{4\,\KK^2}\,\frac{T}{L},
 \ee
 where, following~\cite{dgt}, one notices that $(y^{-3}y')'$ is a total derivative and replaces $y^4$ by $-y_0^4$, thus assuming that the infinite boundary contribution can be dropped. Such a prescription  coincides with normalizing the partition function to the straight line case~\cite{DGO}. The classical contribution to the quark-antiquark potential is obtained dividing the action by the infinite time period $T\to\infty$, thus obtaining $V_{q\bar q}(L)=-\frac{\sqrt{\lambda}\,\pi}{4\,\KK^2\,L}$ as in the leading part of (\ref{classplusoneloop}).

The one-loop correction to the result (\ref{classicalpot}) is obtained by considering fluctuations over the classical solution, a problem addressed in~\cite{Theisen,dgt}. 
Bosonic fluctuations are obtained via a standard background field method,  while  a $\sigma$-dependent rotation in the target space and the standard $\kappa$-symmetry gauge fixing $\theta^1=\theta^2$ for the two Green-Schwarz spinors are used to put the quadratic fermionic term in the Green-Schwarz action into the standard kinetic term for a set of 2-d Majorana fermions. 
In static gauge~\footnote{The  infinite contribution of the ghost determinant is regularized changing the normalization of the non-trivial (longitudinal) bosonic fluctuation, as seen in details in~\cite{dgt}.}  the resulting one-loop partition function is
 \ba\la{Z}
\Gamma_{||}= \frac{\det^{8/2}(-i\,\gamma^\alpha\,\nabla_\alpha+\tau_3)}{\det^{2/2}(-\nabla^2+2)\,\det^{1/2}(-\nabla^2+\textstyle{\frac{1}{4}}R^{(2)}+4)\,\det^{5/2}(- \nabla^2)}~,
\ea
where $R^{(2)}$ is the scalar curvature, $\gamma^0=\tau_2$, $\gamma^1=\tau_1$ and $\gamma_0\gamma_1=-i\,\tau_3$ are the Pauli matrices. As suggested in~\cite{dgt}, it is useful to deal with a conformally flat induced metric, obtained reparametrizing the world-sheet via Jacobi elliptic functions of fixed modulus $k=\textstyle{\frac{1}{\sqrt{2}}}$~\cite{Chu}
\be\la{reparam}
y=\frac{y_0}{\cn\s},~~~~~~~~~~~~t=\frac{\tau}{\sqrt{2}\,y_0}~,
\ee
where now $-\KK<\s<\KK$ and $-\frac{{\cal T}}{2}<\tau<\frac{{\cal T}}{2}$.

The induced metric and the scalar curvature read then
\be\la{induced}
ds^2_{\rm ind}=\frac{1}{2\,\cn^2\s}(d\tau^2+d\s^2),~~~~~~~~~~~~~~~~~
R^{(2)}=-2(1+\cn^4\s)~.
\ee
The explicit expressions for the bosonic differential operators appearing in (\ref{Z}) are then~\cite{Chu}
\ba\la{opbos0}
-\nabla^2&=&-2\cn^2\s\,(\partial_\tau^2+\partial_\s^2)\equiv  2\,\cn^2\s\,{\cal O}_0\\\la{opbos1}
-\nabla^2+2&=&-2\cn^2\s\,(\partial_\tau^2+\partial_\s^2)+2\equiv 2\,\cn^2\s\,{\cal O}_1\\\la{opbos3}
-\nabla^2+4+R^{(2)}&=&-2\cn^2\s\,(\partial_\tau^2+\partial_\s^2)+2(1-\cn^4\s)\equiv 2\,\cn^2\s\,{\cal O}_2
\ea
where the operators ${\cal O}_0$, ${\cal O}_1$ and ${\cal O}_2$ are defined in (\ref{eq})-(\ref{V}) upon Fourier transform of the time variable ($\partial_\tau=-i\,\omega$).  
As suggested in~\cite{Chu}, the fermionic differential operator 
\be\la{opferm}
-i\,\gamma^\alpha\,\nabla_\alpha+\tau_3=\sqrt{2}\,\cn\s\Big[
 -i\Big(\partial_\s+\frac{\sn\s\,\cn\s}{2\,\cn\s}\Big)\,\tau_1-\omega\,\tau_2+\frac{1}{\sqrt{2}\,\cn\s}\tau_3\Big]\equiv\sqrt{2}\,\cn\s\,{\cal O}_\psi
 \ee
can be further diagonalized after squaring it. Using for example $M =\textstyle{\frac{1}{\sqrt{2}}}\big( \begin{smallmatrix} 1&i\\ i&1 \end{smallmatrix}\big)$, one has
\be\la{diag}
{\cal O}^2_\psi=\sqrt{\cn\s}\,M\,{\rm diag}\{{\cal O}_+,\,{\cal O}_-\}\,M^\dagger\frac{1}{\sqrt{\cn\s}}~,
\ee
where ${\cal O}_+$ and ${\cal O}_+$ are defined in (\ref{eq}) and (\ref{V}).
Therefore,  it is $\det^{8/2}{\cal O}^2_\psi\equiv\det^2{\cal O}_+\,\det^2{\cal O}_-$. 

Each ``flat-space'' operator ${\cal O}$ above is rescaled with respect to the original differential operator appearing in (\ref{Z}) via the measure $\textstyle{\frac{1}{\sqrt{g}}}=2\,\cn^2\s$. The finite contribution of such measure to the logarithm of the original determinant (related to the Seeley coefficient which determines the conformal anomaly~\footnote{See discussion in Appendix A of~\cite{dgt}.}) can be explicitly shown to cancel in the ratio (\ref{Z}) of determinants~\cite{Chu}. This justifies the final expression (\ref{Gammapar}) of the effective action.


  \appendix
 \section*{Appendix B:  Relevant elliptic functions and identities }
\refstepcounter{section}
\def\theequation{B.\arabic{equation}}
\setcounter{equation}{0}

The  \emph{incomplete elliptic integrals} of the first and second kind are defined via 
\be\la{incomplete}
F[x,k^2]=\int_0^{x}d\theta\ (1-k^2\,\sin^2\theta)^{-1/2},~~~~~~~~
E[x,k^2] =\int_0^{x}d\theta \ (1-k^2\,\sin^2\theta)^{1/2}
\ee
where $k^2$ is their modulus. The corresponding \emph{complete} elliptic integrals are given by
\be
\KK(k^2)=\KK=F[\textstyle{\frac{\pi}{2}},k^2]\, , ~~
~~~~~~~\EE(k^2)=\EE=E[\textstyle{\frac{\pi}{2}},k^2]\,.
\ee
Defining the \emph{Jacobi amplitude} as
\be
\varphi={\rm am} (u\,|\,k^2),~~~~{\rm where}~~~~u=\int_0^\varphi d\theta\  (1-k'^2\,\sin^2\theta)^{-1/2}
\ee
the \emph{Jacobi elliptic functions} $\sn,\cn,\dn$  are defined by
\be
\sn(u\,|\,k^2)=\sin \varphi,~~~~~~~\cn(u\,|\,k^2)=\cos \varphi,~~~~~~\dn(u\,|\,k^2)=(1-k^2\,\sin^2 \varphi)^{1/2}
\ee
and, for example, ${\rm ns}(u\,|\,k^2)=1/\sn(u\,|\,k^2)$.

Useful relations between the squares of the functions are
\ba
&&-\dn^2(u\,|\,k^2)+k'^2=-k^2\,\cn^2(u\,|\,k^2)=k^2\,\sn^2(u\,|\,k^2)-k^2\\
&&-k'^2\,{\rm nd}(u\,|\,k^2)+k'^2=-k^2\,k'^2\,{\rm sd}^2(u\,|\,k^2)=k^2\,{\rm cd}(u\,|\,k^2)-k^2.
\ea
A useful identity is
\be\la{ellipticF}
\sn^{-1}(z,\textstyle{\frac{1}{2}})=F(\sin^{-1}z,\textstyle{\frac{1}{2}})~.
\ee

The \emph{Jacobi $H$, $\Theta$ and $Z$} functions are defined as follows in terms of the Jacobi $\theta$ functions
\be\label{jacobidef}
H(u\,|\,k^2) = \theta_1\left(\frac{\pi\,u}{2\,\mathbb{K}}, q\right), \qquad
\Theta(u\,|\,k^2) = \theta_4\left(\frac{\pi\,u}{2\,\mathbb{K}}, q\right),\qquad
Z(u\,|\,k^2) = \frac{\pi}{2\,\mathbb{K}}\,\frac{\theta_4'(\frac{\pi\,u}{2\,\mathbb{K}}, q)}
{\theta_4(\frac{\pi\,u}{2\,\mathbb{K}}, q)}
\ee
where $
q=q (k^2) =  \exp(-\pi\frac{\mathbb{K}'}{\mathbb{K}})
$.
A useful identity is
\be\la{zetasplit}
Z(x\,|\,k^2)=E(x\,|\,k^2)-\frac{\EE}{\KK}\,F(x\,|\,k^2)~.
\ee


\appendix
\refstepcounter{section}
\def\theequation{C.\arabic{equation}}
\setcounter{equation}{0}
\section*{Appendix C: \lm problem and determinant via Gelfand-Yaglom method}

Following~\cite{dunne}, consider a Schroedinger  
operator on the interval $x\in [0,L]$ with Dirichlet boundary conditions
\be
\Big[\,-\d^2_x+V(x)\,\Big]\,\psi(x)=\lambda\,\psi(x),~~~~~~~~~~\psi(0)=0,~~~~\psi'(0)=0~.
\ee
Then to compute the determinant one has to solve the associated homogeneous initial value problem
\be\la{initialvalue}
\Big[\,-\d^2_x+V(x)\,\Big]\,\phi(x)=\lambda\,\phi(x),~~~~~~~~~~\phi(0)=0,~~~~\phi'(0)=1
\ee
and
\be\la{detDir}
\det\Big[\,-\d^2_x+V(x)\,\Big]=\phi(L)~~.
\ee
For the single-gap \lm problem
\be\label{lamegen2}
\Big[-\d^2_x+\  2k^2\,\sn^2(x\,|\,k^2)\Big]\,f(x)=\Lambda\,f(x) 
\ee
two independent  solutions   are~\cite{BradenPeriodic}
\be\label{solslame}
f_{\pm}(x) = \frac{H(x\pm\alpha)}{\Theta(x)}\,e^{\mp\,x\, Z(\alpha)} \ ,~~~~~~~
\sn(\alpha\,|\,k^2)= \sqrt{\frac{1+k^2-\Lambda}{k^2}}.
\ee
In terms of them, a solution satisfying the conditions in (\ref{initialvalue}) is 
\be\la{solu}
u(x; \Lambda)&=& \frac{1}{W(\bar x)}\,\Big[f_+(\bar x)\,f_-(x)-f_-(\bar x)\,f_+(x)\Big] 
\ea
where $W$ is the wronskian at a generic initial point $\bar x$
\be\label{wronskian}
W(\bar x)=f_+(\bar x)\,f'_-(\bar x)-f'_+(\bar x)\,f_-(\bar x).
\ee

Exploiting $f_{\pm}(-x)=-f_{\mp}(-x)$ and some properties of the Jacobi elliptic functions it is then easy to check that, in the interval $[-\KK,\KK]$,  the expression for the determinant (\ref{detDir}) yields~\footnote{In the square roots at the second equivalence the known eigenvalues of the \lm equation appear, see for example~\cite{BDFPT}.}
\ba\label{detDirgen}
\!\!\!\!\!\!\!{\rm Det}_{\rm Dir}=u(\KK;\Lambda)= -\frac{\cn\alpha}{\sn\alpha\,\dn\alpha}\,\sinh[2\KK\,Z(\alpha)]=-\frac{\sqrt{1-\Lambda}}{\sqrt{k^2-\Lambda}\,\sqrt{1+k^2-\Lambda}}\,\sinh[2\KK\,Z(\alpha)]~.
\ea
The determinants (\ref{detO1})-(\ref{detOf}) evaluated in Section 2 are  generalizations of the expression (\ref{detDirgen}). Their slightly more involved form is simply due to the presence of the infrared regulator $\epsilon$, which alters the boundary conditions of the problem.

\appendix
\refstepcounter{section}
\def\theequation{D.\arabic{equation}}
\setcounter{equation}{0}
\section*{Appendix D: Equivalent form of the integral $\cal I$ in (\ref{integral})}

Basic manipulation of the special functions in (\ref{integral}) with identities such as (\ref{ellipticF}) and (\ref{zetasplit})  leads to the following expression 
\ba\la{integral2}
\!\!{\cal I}=\!\!\!\int_{0}^{k}\!\!\! d\omega \,\ln\Big[\frac{\cosh^8 \tilde x_f}{\sin^2 \tilde x_{1{\rm zero}}\,\sinh \tilde x_2\,\sinh^5[2\,\KK\,\o]}\Big]+\int_{k}^{\infty}\!\!\!d\omega \,\ln\Big[\frac{\cosh^8 \tilde x_f}{\sinh^2 \tilde x_{1{\rm inf}}\,\sinh \tilde x _2\,\sinh^5[2\,\KK\,\o]}\Big] ~,
\ea 
where
\ba
\tilde x_f&=&\textstyle{\frac{\pi \, \omega}{\KK}} \,{}_2F_1\left[\textstyle{\frac{1}{2}},\textstyle{\frac{1}{4}},\textstyle{\frac{5}{4}};-16\, \omega ^4\right]+\frac{8 \,\omega ^3 \,\KK}{3}\,{}_2F_1\left[\textstyle{\frac{1}{2}},\textstyle{\frac{3}{4}},\textstyle{\frac{7}{4}};-16\, \omega ^4\right]~,\\\no
\tilde x_{1{\rm zero}}&=&-2\,  \omega\,\KK+4\,\omega \,\EE\,\,  {}_2F_1\left[\textstyle{\frac{1}{2}},\textstyle{\frac{1}{4}},\textstyle{\frac{5}{4}};4 \omega ^4\right]+\textstyle{\frac{4 \,\KK \,\omega ^3}{3}}\,{}_2F_1 \left[\textstyle{\frac{1}{2}},\textstyle{\frac{3}{4}},\textstyle{\frac{7}{4}};4
   \omega ^4\right]+\\
   &&-\textstyle{\frac{8\, 2^{1/4}\,\KK\, \omega ^5}{5\, (1+\sqrt{1-4 \omega ^4})^{5/4}}}\,{}_2F_1\left[\textstyle{\frac{3}{4}},\textstyle{\frac{5}{4}},\textstyle{\frac{9}{4}};\textstyle{\frac{1}{2}
   (1-\sqrt{1-4 \omega ^4})}\right]~,\\\no
   \tilde x_{1{\rm inf}}&=&\textstyle{\frac{\KK}{\omega}}(1+ \sqrt{4 \omega
   ^4-1})-\frac{2 \,\EE}{\omega}\,{}_2F_1 \left[\textstyle{\frac{1}{4}},\textstyle{\frac{1}{2}},\textstyle{\frac{5}{4}};\frac{1}{4 \omega ^4}\right]+\frac{\KK}{6\,
   \omega ^3}\,{}_2F_1 \left[\textstyle{\frac{1}{2}},\textstyle{\frac{3}{4}},\textstyle{\frac{7}{4}};\frac{1}{4 \omega ^4}\right]+\\
   &&+\textstyle{\frac{\KK\, (2 \omega ^2+\sqrt{4 \omega ^4-1})^{3/4} }{5 \,\sqrt{2} \,\omega ^{5/2} \,(8 \omega ^4-1+4 \sqrt{4 \omega ^4-1} \omega ^2)}}\,\,{}_2F_1 \left[\textstyle{\frac{3}{4}},\textstyle{\frac{5}{4}},\textstyle{\frac{9}{4}};\textstyle{\frac{1}{2}}-\textstyle{\frac{\sqrt{4 \omega ^4-1}}{4 \omega
   ^2}}\right]~,\\\la{xtilde2}
   \tilde x_2&=&\textstyle{\frac{\pi \, \omega}{\KK}}  \,{}_2F_1\left[\textstyle{\frac{1}{2}},\textstyle{\frac{1}{4}},\textstyle{\frac{5}{4}};-\omega ^4\right]+\frac{2 \,\omega^3\,\KK}{3} \,{}_2F_1\left[\textstyle{\frac{1}{2}},\textstyle{\frac{3}{4}},\textstyle{\frac{7}{4}};-\omega ^4\right]~.
\ea

A compact way to write the numerical contribution ${\cal I}^{\rm num}$ in (\ref{anandnum}) is obtained  as follows.  Each of the  (\ref{xf})-(\ref{x2}) can be put in a simpler form with   the change of variables $\o=\frac{1}{2}\tan\frac{\a}{2}$ in (\ref{xf}), $\o=\frac{1}{\sqrt{2}}\cos\frac{\a}{2}$ in (\ref{x1zero}),  $\o=\frac{1}{\sqrt{2}}\sec\frac{\a}{2}$ in (\ref{x1inf}) and $\o=\tan\frac{\a}{2}$ in (\ref{x2}). One obtains
\ba
\bar x_f&=&\frac{1}{2}\Big(\frac{\pi}{2\,\KK}+\KK\Big)\,F[\alpha]-\KK\,E[\a]+\frac{\KK}{2}\,\sin\a\,\sqrt{1+\tan^4\frac{\a}{2}}\\
\bar x_{1{\rm zero}}&=&\Big(\frac{\pi}{2\,\KK}-\KK\Big)\,F[\textstyle{\frac{\alpha}{2}}]+2\,\KK\,E[\textstyle{\frac{\alpha}{2}}] \\
\bar x_{1{\rm inf}}&=&\Big(\frac{\pi}{2\,\KK}+\KK\Big)\,F[\textstyle{\frac{\alpha}{2}}]-2\KK\,E[\textstyle{\frac{\alpha}{2}}]+ \KK \tan\frac{\alpha}{2}\, \sqrt{3 + \cos\a} \\
\bar x_2&=&\ \Big(\frac{\pi}{2\,\KK}+\KK\Big)\,F[\alpha]-2\,\KK\,E[\a]+\KK\,\sin\a\,\sqrt{1+\tan^4\frac{\a}{2}}~.
\ea
which makes explicit  $x_2=2x_f$.
In terms of the variables above, the contributions (\ref{numf})-(\ref{num2}) are summed as~\footnote{We used Mathematica numerical integration here with, for example, 
${\rm WorkingPrecision}\to100$, ${\rm PrecisionGoal}\to 100$, ${\rm MaxRecursion}\to 30$. We have only reported in (\ref{numerics}) the first 50 digits.}
\be\no
{\cal I}^{\rm num}\!\!\!\!\!&=&\!\!\!\!\!\int_0^{\pi} \frac{d\a}{2\,\cos^2\textstyle{\frac{\alpha}{2}}}
\Big[\ln\frac{(1+e^{-2 \bar x_f})^3}{1-e^{-2\,\bar x_f}}-\sqrt{2} \,\sin\textstyle{\frac{\a}{2}}
\Big(\ln(1-e^{-2\,x_{1{\rm inf}}})+\cos^2\frac{\a}{2}\,\ln\sin x_{1{\rm zero}}\Big)\Big]\\\la{numerics}
&=&3.09111054729005989778296487945453992761532660548813~.
\ee




\end{document}